\documentclass{article} 
\usepackage[top=1.2in, bottom=1.2in, left=1.2in, right=1.2in]{geometry}

\usepackage{amsmath,amssymb,amsfonts,amsthm}
\allowdisplaybreaks[1] 

\usepackage{color}
\usepackage[usenames,dvipsnames]{xcolor}

\usepackage{cite}	
\usepackage{url}

\theoremstyle{definition}

\theoremstyle{definition}

\usepackage{hyperref}
\hypersetup{
   colorlinks=true,
   citecolor=Green
}

\usepackage{algorithmic}
\usepackage[ruled, noend,lined,linesnumbered]{algorithm2e} 

\newcommand\domain{\mathcal{X}}
\newcommand\report{CRT report}

\begin{document}
\title{On the Privacy of the Opal Data Release: A Response}

\author{Hassan Jameel Asghar, Paul Tyler and Mohamed Ali Kaafar\\\\
\small Data61, CSIRO, Australia\\
\small \{\texttt{hassan.asghar, paul.tyler, dali.kaafar}\}\texttt{@data61.csiro.au}
}

\date{\today}

\maketitle
\begin{abstract}
This document is a response to a report from the University of Melbourne on the privacy of the Opal dataset release. The Opal dataset was released by Data61 (CSIRO) in conjunction with the Transport for New South Wales (TfNSW). The data consists of two separate weeks of ``tap-on/tap-off'' data of individuals who used any of the four different modes of public transport from TfNSW: buses, light rail, train and ferries. These taps are recorded through the smart ticketing system, known as Opal, available in the state of New South Wales, Australia. 
\end{abstract}

\section{Introduction}

This document is in relation to a report from Culnane, Rubinstein and Teaque~\cite{melb-uni} assessing the privacy of the Opal dataset release from Data61 (CSIRO) in conjunction with the Transport for New South Wales (TfNSW)~\cite{opal-data, opal-tech-report}. Briefly, the Opal dataset is a set of trips from commuters using different public transport modes available in the state of New South Wales (NSW), Australia. These trips were recorded through the smart ticketing system, known as Opal, available in NSW. A sample of this data was sanitized via a differentially private treatment by Data61~\cite{opal-tech-report} and TfNSW made the resulting datasets available for download as open data~\cite{opal-data}. Details about the Opal dataset and its privacy preserving release can be found in~\cite{opal-tech-report, opal-data}. The purpose of this document is to address some points raised by Culnane et al. in~\cite{melb-uni} regarding the privacy of the release. We shall refer to this as the {\report}. 
It should be noted that our technical report~\cite{opal-tech-report} was not available to Culnane et al. 

\section{Response}
\label{sec:resp}
In what follows, we addressed some of the points raised in the {\report}. We recommend the reader to read our technical report~\cite{opal-tech-report} before reading this response. 

\subsubsection*{Aggregation of Stops, Routes and Times}
It is hinted in the {\report} that the aggregation of some stops (tap-on or off locations) and routes (trips) is done to protect privacy. In actual, as discussed in~\cite[\S 4.1]{opal-tech-report}, the stops and times were aggregated and binned, respectively, to ensure that the dataset is sufficiently dense.\footnote{See Definition 7 in~\cite[\S 2]{opal-tech-report} for what we mean by dense versus sparse datasets.} This in turn ensures that sufficiently many points are returned by the SBH algorithm.\footnote{The SBH (Stability-based Histogram) algorithm appears in~\cite[\S 7.1]{salil-tut},\cite{balcer} which slightly modifies the online version of the algorithm proposed in~\cite{bun-stable}. The algorithm is reproduced in~\cite[\S 3]{opal-tech-report}.} For instance, the number of stops for buses were far too many to result in a dense dataset. Likewise the original times, that were within a one minute interval, would have resulted in a sparse dataset, resulting in not many points released in the output through the SBH algorithm. Thus, aggregation of stops and times was done to make the domain narrower and hence the datasets more dense. This was done from a utility perspective, and not for privacy.  Privacy follows automatically from the differential privacy of the process. 

\subsubsection*{Decoupling of Tap-on and Tap-off Events}
The decoupling of these events was done once again to make the two separate datasets more dense, and hence to enable the algorithm to output more points. Thus, this was done not to protect privacy as is indicated in the {\report}, but to improve utility. Indeed, one could use the algorithm with both the tap-on/tap-off events, i.e., each row containing tap-on and off times and locations. The only problem would be that the resulting dataset would be more sparse, and hence many points from the original dataset would be removed in the synthetic output. The issue would not be privacy but the utility of the resulting dataset.

\subsubsection*{Detection of Individuals or Groups}
The {\report} states that the algorithm used to release the Opal dataset is at best $(\epsilon, \delta)$-differentially private with a small but constant $\delta$, as opposed to $\delta$ being negligible in the size of the dataset. As mentioned in~\cite[\S 4]{opal-tech-report}, the released Opal dataset is indeed $(\epsilon, \delta)$-differentially private. For $\delta$, it is true that in theory it is supposed to be negligible in the size of the dataset. More precisely, it should be smaller than any inverse polynomial power of $n$, where $n$ is the size of the dataset $D$~\cite{dp-book, salil-tut}. However, in practice, we need to instantiate $\delta$ with some value. Once we instantiate $\delta$, by definition it is a constant, and no longer equivalent to its theoretical definition. Therefore, we need to ensure that $\delta$ is a small enough constant so that the event associated with $\delta$ is unlikely to occur in practice. By looking at the SBH algorithm, we see that we need to set $\delta$ such that the probability of the event that a trip that is only performed by a single individual, i.e., the \emph{bad event} defined in the proof of Theorem~5~\cite[\S 3.2]{opal-tech-report}, is small. This probability is given by $\delta_j/2$, where $j$ denotes one of the six column-wise partitions. We fixed $\delta_j = 1/8000000 \approx 2^{-23}$, making this probability approximately $2^{-24}$. The value of $\delta_j \approx 2^{-23}$ implies that the overall value of $\delta$ (by basic composition) is less than $2^{-20}$ or $10^{-6}$. This value of $\delta$ has been used before in the literature~\cite[\S 6]{dual-query}, and is slightly far from another value of $2^{-30}$ recommended in~\cite[\S 8, p. 11]{sec-sample}. Note that together with $\epsilon = 2$, this meant that the threshold of the SBH algorithm for the two-column datasets was 18. This number protects the aforementioned bad event for individuals as well as small groups (of say, size 5) with high probability. A much smaller value of $\delta_j$ and subsequently $\delta$ could be chosen. However, smaller values result in a higher threshold in the SBH algorithm, producing less points (trips) in the output. This was a compromise made between utility and privacy. 

\subsubsection*{Pure versus Approximate Differential Privacy}
The {\report} states that pure differential privacy should be preferred over approximate differential privacy since the latter is a weaker notion of privacy. There are varying opinions around the merits of approximate differential privacy as opposed to pure differential privacy. Pure differential privacy is indeed a stronger notion of privacy, where as approximate differential privacy is a \emph{slight} relaxation. The relaxation allows to sometimes develop more efficient algorithms with more accurate answers to specific queries than what is allowed by pure differential privacy~\cite{dp-book, salil-tut}. Approximate differential privacy essentially means that the output is pure differentially private except for a small probability characterised by $\delta$. This probability should be small enough for such an event to be improbable, hence a value of $\delta$ negligible in the size of the dataset is recommended. This is similar to an analogous practice in cryptography, where the success probability of an adversary attempting to break the cryptosystem is required to be negligible in the size of the security parameter. Thus, while pure differential privacy is preferable, in the absence of efficient and high utility algorithms, we do not believe there is any harm in using approximate differential privacy as long as it is understood what privacy implications may occur if the value of $\delta$ is not chosen appropriately.

\subsubsection*{Use of the Algorithm on Sensitive Data}
It is mentioned in the {\report} that our approach should not be used on more sensitive datasets for privacy reasons. However, as long as the notion of approximate differential privacy is considered an acceptable definition of privacy, there is no distinction between sensitive or non-sensitive data. Indeed the algorithm by definition ensures that sensitive data points are almost always never output (provided appropriate $\epsilon$ and $\delta$ are chosen)! Sensitive data by nature will make each row unique and will result in the input dataset being sparse.\footnote{There might be some confusion in exactly what is meant by sensitive data. Consider a dataset that contains your full name. This by nature ``sounds'' like sensitive data. But if many other people in the dataset share the same full name as you, the data is no longer sensitive. The data is sensitive only if it is rare or unique to you. In other words, if it is only learnable through your data being in the dataset (or through the data of a small group being in the dataset).}  It is the construction of the SBH algorithm that unique points are not released (with overwhelming probability, based on the value of $\delta$). The algorithm protects privacy regardless of the nature of the data, as long as an appropriate value of $\delta$ is chosen. Thus, there is nothing inherently wrong with our approach. The only risk is that of poor utility in the released dataset, if it predominantly consists of sensitive (highly unique) points.

\subsubsection*{Differential Privacy versus Perfect Privacy}
The {\report} states that there is a difference between perfect privacy and differential privacy, and the two are not the same. This is true in principle, but can be misleading in making the reader believe that there are some privacy vulnerabilities surrounding differential privacy. Perfect privacy can be defined as the posterior and prior knowledge of an adversary remaining the same after it has observed the output of an algorithm over some dataset. Within the differential privacy framework, by choosing an $\epsilon = 0$, it is possible to ensure that the probability of observing the same output from two neighbouring datasets is exactly the same, thus achieving perfect privacy. This means, however, that we gain nothing useful from the dataset~\cite[\S 4.1, p. 78]{exposed}. What we observe from the dataset could easily have been observed with some random output generated by simply ignoring the input dataset. Differential privacy protects privacy of individuals while enabling statistical insights about the population comprising the dataset. This is done by making the outputs indistinguishable between neighbouring datasets (or indistinguishable with high probability in case of approximate differential privacy). From the privacy point of view, this notion relates to the privacy of individuals as opposed to all individuals within the dataset taken together. Insights on populations is what we seek to learn from the dataset.

\subsubsection*{Public Disclosure of Differentially Private Algorithms and Privacy Parameters}
Here we agree with the conclusion of the {\report} that any algorithm purported to provide differential privacy should be disclosed together with a proof of privacy and the parameters used to generate data. It is also illuminating to see how the values of the privacy parameters are inferred in the {\report} from the released dataset. Notice that these parameters are not required to be kept private according to the definition of differential privacy. The full description of the process used to produce the Opal dataset as well as the parameters used can be found in~\cite{opal-tech-report}. However, it should also be noted that our engagement was also of a commercial nature and commercial concerns can impact what can and cannot be published. 

\subsubsection*{To Perturb or Not to Perturb Zeroes}
\label{sub:zeroes}
The {\report} recommends the use of an alternate algorithm that would also perturb points that have a value of $0$ in the input dataset. That is, even those points that are absent from the input dataset but are otherwise points within the data domain. From the Opal dataset point of view, this means even adding noise to trips that are absent in the input dataset. There are a couple of issues with this approach which we highlight below.

Developing differentially private algorithms to produce synthetic datasets is in general a difficult task. This is somewhat evident from the definition of differential privacy. Recall that in order for an algorithm to be differentially private we need to consider \emph{all possible} datasets $D \in \domain^n$ and their neighbours. If an algorithm uses as input all possible points in $\domain$, depending on how big $\domain$ is, it may become computationally infeasible. For instance, a straightforward mechanism to output the histogram of some dataset $D$, i.e., answering all point functions, is to apply Laplace noise of magnitude $1/\epsilon$ to the answers $a_x(D)$ to each point $x \in \domain$ (even points that are not present in $D$). However, in general $\domain$ can be too large for this to be a practical solution. Consider a domain $\domain$ having 4 attributes, each with range of magnitude $1000$. Then the size of the domain, i.e., $| \domain |$ is $10^{12}$. Thus, the above algorithm will take time approximately proportional to $2^{40}$ to release a synthetic dataset. This means that adding noise to all $0$ counts will result in an algorithm that is infeasible with a large domain size.

Another issue with this approach is that the algorithm will then output points, i.e., trips, that have no meaning in the real world. For instance, from the Opal dataset point of view, it might output a trip whose route does not exist. Of course, in a post-processing step, these anomalies can be ruled out, but doing so in general is not easy, as one needs to check each row for consistency. In the absence of rules to check for consistency in an automated way, this becomes a prohibitive task. Such anomalous trips could indeed be left as is in the output dataset. But this poses a problem from a usability point of view. An analyst has no idea (without reference to a detailed route and time-table) whether a given trip is indeed a valid trip or a by product of the synthetic data generation process. The problem is exacerbated by noting that due to the nature of transport there could be delays in services resulting in the dataset showing trips far off from their expected schedule in the time table. Similarly, issues such as cancellation of services or other external factors may be significant. Nevertheless, the above mentioned algorithm could be used if a dictionary of all possible trips (tap locations and time combinations) is provided, and such a list is not too large from a computational point of view. Due to time constraints, this approach was ruled out for the Opal data release but could be considered for any future releases.

\subsubsection*{Creating Different Synthetic Datasets from the Same Dataset}
The {\report} states that constructing different datasets from the same input datasets can have unexpected consequences. This alludes to the 6 different synthetic datasets produced from the same input dataset. It is further recommended that this be done by first producing a differentially private version of the original dataset (taking all tap locations and times together) and then constructing individual datasets from the result. We chose not to apply the SBH algorithm on the complete trip as it made the original dataset too sparse, and hence the output would omit many points; an undesirable outcome from a utility point of view. Thus, we decided to split the data and apply the algorithm separately on each part. However, we do not agree that obtaining several datasets from the same input dataset has unexpected consequences. The consequences are those that can be expected from the privacy parameters of the overall release. We have done this by showing that the overall output is $(\epsilon, \delta)$-differentially private. Thus, each of the datasets from the same original dataset may be $(\epsilon_j, \delta_j)$-differentially private, with a smaller $\epsilon_j$ (and $\delta_j$), the privacy guarantee is only in terms of the overall $\epsilon$ and $\delta$. Thus, the consequences are those that could result when this higher $\epsilon$ is taken into account (and hence not unexpected). 

The {\report} gives a few examples of how releasing several synthetic datasets from the same original dataset can result in guessing the frequency of trips that were omitted due to a frequency less than the threshold in the SBH algorithm. But this is not unexpected and can be explained through the help of a simpler example. Suppose a dataset has only one column named `Gender.' Further suppose the dataset contains $n = 110$ rows, 100 of them male and 10 of them female. Suppose now we use the SBH algorithm with parameters $\epsilon = 1$ and $\delta = 2^{-24}$. This means that the threshold of the SBH algorithm is approximately 35. We now run the SBH algorithm twice: first on the counts of males and females, and secondly on the total count $n = 110$.\footnote{The total privacy budget is therefore $\epsilon = 2$, by basic composition.} With overwhelming probability, the output will have the number of males around $100$, say 101, the number of females $0$, and the total count as approximately $110$, say 109. From this, we can deduce that the total number of females in the dataset was around 8; close to the value 10 which the algorithm tried to suppress. Does this constitute a privacy violation? No. Recall the reason behind the threshold check. Intuitively, the algorithm is trying to hide points that do not exist in the original dataset behind points that have a low count, thus making the two indistinguishable. In this example, there are no other points in the domain (the domain is simply male or female gender). Thus, the algorithm might as well have output the (noisy) number of females, without any privacy violation. This will be $(\epsilon, 0)$-differentially private.\footnote{Notice also that since we access the dataset twice, first for number of males and females and then for $n$, the mechanism is $(2 \epsilon, 2 \delta)$-differentially private. Therefore, the resulting threshold would also be lower (around 18) if we had only chosen to answer the number of males and females with this higher privacy budget. The resulting threshold is still higher than the number of females.} The examples given in the {\report} highlight precisely these cases. For instance, the example of the ferry trips given in Section~4.1 of the {\report} indicates that the Cremorne Point wharf was the only possible trip within that time period. The algorithm could have simply output the noisy count of people at that point without any privacy violation.  But by construction, the algorithm does not do so. 

Thus, deriving several datasets from the same original dataset does not constitute a privacy violation as long as it is ensured that the collection is altogether $(\epsilon, \delta)$-private with a precisely quantified (bounded) $\epsilon$ and $\delta$ via the composition theorem. As this is indeed the case with the different synthetic datasets of the Opal data, this does not constitute a privacy breach. This should also be clear from online mechanisms used in differential privacy, which allow multiple queries on the same dataset, even repeat queries, which is acknowledged in the \report.

\subsubsection*{Ruling out Datasets with Non-Zero Counts for Trips}
The {\report} states that not perturbing the zero counts is an error, which could be used to rule out certain datasets from the set of possible input datasets (given to the SBH algorithm). However, note that the definition of differential privacy is around providing privacy for individuals, not for entire datasets. By choosing $\epsilon$ and $\delta$, it can be ensured that the threshold is set to a level which hides the zero counts behind other low counts, thus providing protection for trips. Differential privacy does not apply to protection of privacy for entire datasets. The definition considers all possible datasets and its neighbours; not all possible datasets and all its \emph{distant} neighbours. Such a definition (which again, is not differential privacy) will provide limited utility and may hide insights into statistical trends from datasets.

\subsubsection*{Correlated Rows}
Some recent works have pointed out a supposed flaw in the definition of differential privacy namely that it has an assumption around the rows of a dataset being independent~\cite{free-lunch, ndss-dependence}. The {\report} also mentions this and states that when the assumption fails to hold, then it reduces the privacy guarantees (less than what a given value of $\epsilon$ suggests). However, this critique of differential privacy is incorrect. By looking at the definition of differential privacy we see that no assumption on the independence or dependence of rows is made. The definition only seeks to define privacy as the indistinguishably in the output after exclusion of the individual from the dataset. If many individuals happen to have correlated data, then it is precisely what the analyst seeks to learn from the data: the general trend~\cite{exposed}. An oft-repeated example in the differential privacy literature is that of ``smoking causes cancer.'' A dataset containing a list of smokers and a column indicating presence/absence of cancer is expected to have many correlated rows. Differential privacy does not guarantee that removal of an individual from the dataset will hide the correlation between smoking and cancer. Once the correlation is learned, then knowing that the target individual smokes reveals sensitive information about the individual (high chance of cancer). If the population-wise correlations are to be hidden, then there is nothing useful that can be learned from the data, which defeats the purpose of releasing the data in the first place. See further discussion on this distinction in~\cite[\S 1.1.4]{exposed}.\footnote{Also see the less formal but highly informative blog posts by Frank McSherry titled ``Lunchtime for Data Privacy'' and ``Differential privacy and correlated data'' at \url{https://github.com/frankmcsherry/blog}.}

Having said that, the notion of privacy used by us is that of \emph{trip privacy}. That is, each row of the dataset $D$ represents a single trip from some individual. That is, in the definition of differential privacy two neighbouring datasets differ in one trip. An individual may have multiple trips in the dataset. Thus, in this sense we ensure that the addition or removal of a single trip of an individual does not change the output of the differentially private algorithm beyond what is allowed by the privacy parameter $\epsilon$. This is akin to providing edge privacy as opposed to node privacy in the context of graph data~\cite[\S 1]{graph-privacy}; the latter being a stronger notion of privacy. We in fact first partition the Opal dataset according to date and transport modes, so that for each transport mode and date pair, we obtain a separate partition. This then amounts to providing privacy of a trip of an individual on a particular date and transport mode. Since any identifying information has already been removed in the output datasets, and since the datasets have been decoupled into different datasets based on transport modes and dates, there is no obvious link between the trips. Therefore, we believe that the somewhat weaker notion of trip privacy, as opposed to all trips from an individual, is reasonable in this case. Using the stronger notion of privacy means that addition/removal of an individual may effect multiple rows of the dataset (say trips made on the same day and same transport mode, or trips made on different days with different transport modes). It is then not easy to obtain a tight upper bound on the sensitivity that would result in reasonable perturbation through the Laplace mechanism from a utility point-of-view.


\bibliographystyle{unsrt}
\bibliography{response}

\begin{thebibliography}{10}

\bibitem{melb-uni}
Chris Culnane, Benjamin I.~P. Rubinstein, and Vanessa Teague.
\newblock {Privacy Assessment of De-identified Opal Data: A report for
  Transport for NSW}.
\newblock {\em CoRR}, abs/1704.08547, 2017.

\bibitem{opal-data}
Transport for NSW.
\newblock {Opal Tap On and Tap Off}.
\newblock
  \url{https://opendata.transport.nsw.gov.au/dataset/opal-tap-on-and-tap-off},
  2017.
\newblock Accessed 10 May, 2017.

\bibitem{opal-tech-report}
Hassan~Jameel Asghar, Paul Tyler, and Mohamed~Ali Kaafar.
\newblock {Differentially Private Release of Public Transport Data: The Opal
  Use Case}.
\newblock {\em CoRR}, abs/1705.05957, 2017.

\bibitem{salil-tut}
Salil Vadhan.
\newblock {The Complexity of Differential Privacy}.
\newblock Available online
  \url{http://privacytools.seas.harvard.edu/files/privacytools/files/complexityprivacy_1.pdf},
  2016.

\bibitem{balcer}
Victor Balcer and Salil Vadhan.
\newblock {Efficient algorithms for differentially private histograms with
  worst-case accuracy over large domains}.
\newblock Manuscript, 2016.

\bibitem{bun-stable}
Mark Bun, Kobbi Nissim, and Uri Stemmer.
\newblock Simultaneous private learning of multiple concepts.
\newblock In {\em Proceedings of the 2016 ACM Conference on Innovations in
  Theoretical Computer Science}, ITCS '16, pages 369--380, New York, NY, USA,
  2016. ACM.

\bibitem{dp-book}
Cynthia Dwork and Aaron Roth.
\newblock {The Algorithmic Foundations of Differential Privacy}.
\newblock {\em Foundations and Trends in Theoretical Computer Science},
  9(3â4):211--407, 2014.

\bibitem{dual-query}
Marco Gaboardi, Emilio JesÃºs~Gallego Arias, Justin Hsu, Aaron Roth, and
  Zhiwei~Steven Wu.
\newblock {Dual Query: Practical Private Query Release for High Dimensional
  Data}.
\newblock In {\em {Proceedings of the 31st International Conference on Machine
  Learning, Cycle 2}}, volume~32 of {\em {JMLR Proceedings}}, pages 1170--1178.
  {JMLR.org}, 2014.

\bibitem{sec-sample}
Marco Gaboardi, James Honaker, Gary King, Kobbi Nissim, Jonathan Ullman, and
  Salil~P. Vadhan.
\newblock {PSI} ({\(\Psi\)}): a private data sharing interface.
\newblock {\em CoRR}, abs/1609.04340, 2016.

\bibitem{exposed}
Cynthia Dwork, Adam Smith, Thomas Steinke, and Jonathan Ullman.
\newblock {Exposed! A Survey of Attacks on Private Data}.
\newblock {\em Annual Review of Statistics and Its Application}, 4(1):61--84,
  2017.

\bibitem{free-lunch}
Daniel Kifer and Ashwin Machanavajjhala.
\newblock No free lunch in data privacy.
\newblock In {\em Proceedings of the 2011 ACM SIGMOD International Conference
  on Management of Data}, SIGMOD '11, pages 193--204, New York, NY, USA, 2011.
  ACM.

\bibitem{ndss-dependence}
Changchang Liu, Supriyo Chakraborty, and Prateek Mittal.
\newblock Dependence makes you vulnerable: Differential privacy under dependent
  tuples.
\newblock In Srdjan Capkun, editor, {\em NDSS 2016, Network and Distributed
  System Security Symposium}, Geneva, Switzerland, February 2016. Internet
  Society, Internet Society.

\bibitem{graph-privacy}
Shiva~Prasad Kasiviswanathan, Kobbi Nissim, Sofya Raskhodnikova, and Adam
  Smith.
\newblock Analyzing graphs with node differential privacy.
\newblock In {\em Proceedings of the 10th Theory of Cryptography Conference on
  Theory of Cryptography}, TCC'13, pages 457--476, Berlin, Heidelberg, 2013.
  Springer-Verlag.

\end{thebibliography}

\end{document}